\documentclass[twocolumn,prl,showpacs]{revtex4}
\begin{document}

\title{\bf Environment - Assisted Invariance, Causality, and Probabilities
in Quantum Physics}

\author{Wojciech Hubert Zurek}
\address{Theory Division, MS B288, LANL, Los Alamos, New Mexico 87545}


\begin{abstract}

I introduce environment - assisted invariance -- a symmetry
related to causality that is exhibited by correlated quantum states -- and 
describe how it can be used to understand the nature of ignorance and, hence, 
the origin of probabilities in quantum physics.

\end{abstract}

\maketitle

\smallskip

Quantum theory has a peculiar feature conspicuously absent from classical 
physics: One can know precisely the state of a composite object (consisting, 
for example, of the system ${\cal S}$ and the environment ${\cal E}$) and 
yet be ignorant of the state of ${\cal S}$ alone. The purpose
of this paper is to introduce environment - assisted invariance, or 
{\it envariance} to capture this counterintuitive quantum symmetry that allows 
observer to use his prefect knowledge (of ${\cal SE}$) as a proof of his 
ignorance of ${\cal S}$: When a $u_{\cal S}$ acting on ${\cal S}$ alone 
can be undone by a transformation acting solely on ${\cal E}$, so that 
the joint state of ${\cal SE}$ is unchanged, this state will be said to be 
envariant with respect to $u_{\cal S}$. 

Clearly, envariant properties do no belong ${\cal S}$ alone. 
Hence, entanglement between ${\cal S}$ and ${\cal E}$ that enables 
envariance implies ignorance about ${\cal S}$. Envariance is 
associated with phases of the Schmidt decomposition of the state 
representing ${\cal SE}$. It anticipates some of the consequences 
of environment - induced superselection (or ``einselection'') and allows one 
to derive and interpret Born's rule$^1$ relating quantum amplitudes and 
probabilities through an appeal to causality, in a manner that is different
and more physically motivated than the famous theorem of Gleason$^2$. 
Thus envariance leads to one of the central elements of the interpretation 
of quantum theory from first principles. 

It has become increasingly popular to associate the transition from 
quantum to classical with decoherence$^{3-5}$ and its key consequence,
einselection of the preferred set of pointer states$^{6,7}$. Pointer states 
remain unperturbed in spite of immersion of the system in the 
environment. This allows for predictability and other symptoms of 
`objective existence', cornerstones of classicality.
However, while this line of reasoning has had notable successes, its very
foundation is sometimes regarded as {\it ad hoc}, opening it to a charge of
providing the solution ``for practical purposes only''$^8$. In particular,
it was pointed out by supporters and detractors alike$^{9-11}$
that the relation between the quantum states and probabilities 
is {\it not} settled by decoherence: Born's rule has to be postulated 
separately. Hence, as it is used to arrive at the concept of reduced density 
matrix$^{12,13}$ -- the key tool of the decoherence program -- derivations
of Born's rule based on decoherence would be suspect on grounds of 
circularity.

The physical
motivation of envariance is simple: We imagine a system ${\cal S}$ entangled
with a distant (or, at least, dynamically decoupled) environment ${\cal E}$.
The question we now pose is: Given the state of the combined ${\cal SE}$
expressed in the Schmidt form:
$$ |\psi_{\cal SE} \rangle \ = \
\sum_{k=1}^N \alpha_k |\sigma_k \rangle | \varepsilon_k \rangle \ ,  \eqno(1)$$
with complex $\alpha_k$ and with $\{|\sigma_k\rangle \}$ and
$\{|\varepsilon_k \rangle\}$ orthonormal,  what can one know about ${\cal S}$?

The usual answer would be to use $\psi_{\cal SE}$ to obtain reduced 
density matrix of the system:
$$ \rho_{\cal S} \ = \ Tr_{\cal E}|\psi_{\cal SE}\rangle\langle\psi_{\cal SE}|\
= \  \sum_{k=1}^N |\alpha_k|^2 |\sigma_k\rangle\langle \sigma_k| \eqno(2)$$
This step presumes Born's rule ($p_k=|\alpha_k|^2$ is employed$^{12,13}$ to 
get from Eq. (1) to (2)). Therefore, we cannot take it: We are looking for 
a more fundamental reason to {\it trace out the environment}, and we aim to 
{\it derive} Born's rule. If succesfull, such derivation would in turn 
justify tracing, reduced density matrices, etc.

In order to proceed in this fundamental direction we can rely only on these
principles of quantum theory that manifestly do not employ Born's rule.
To this end, we identify properties of the entangled state $\psi_{\cal SE}$
that do not belong to ${\cal S}$ alone. The strategy is straightforward:
Apply transformations that act on the Hilbert space ${\cal H_S}$ of the system 
and investigate whether their effect {\it on the joint state} $\psi_{\cal SE}$
can be undone by the ``countertransformations'' acting solely on ${\cal H_E}$.
When the transformed property of the system can be so ``untransformed'' by
acting solely on the environment, it is not the property of ${\cal S}$.
Hence, when ${\cal SE}$ is in the state $|\psi_{\cal SE}\rangle$ with this
characteristic, it follows that the envariant properties of ${\cal S}$ must
be completely unknown.

This motivating discussion leads to the definition of envariance:
When for a certain $|\psi_{\cal SE} \rangle $ and for a transformation
$U_{\cal S} = u_{\cal S} \otimes {\bf 1}_{\cal E}$ there exists
a corresponding $U_{\cal E} = {\bf 1}_{\cal S} \otimes u_{\cal E}$ such that:
$$ U_{\cal S} U_{\cal E} |\psi_{\cal SE} \rangle \ = |\psi_{\cal SE}
\rangle \ ,
\eqno(3)$$
then for this state the properties of ${\cal S}$ affected -- transformed --
by $u_{\cal S}$ (and, in particular, connected with any observables that
do not commute with $u_{\cal S}$) are {\it envariant}.

To paraphrase Bohr's famous dictum about quantum theory, ``if the reader does
not find envariance of pure quantum states strange, he has not understood
it'': A state of two  coins (say, a penny and a cent) would be envariant
when first the penny could be flipped, and then the cent ``counterflipped'' 
in a manner that leaves their joint state unchanged, i.e., when there 
measurement on ${\cal SE}$ will confirm that it is still in the same joint
state. Pure (perfectly known) quantum states can exhibit this 
symmetry, as we shall see below. If this was ever true classically, one would 
conclude that the observer could not tell the difference caused by the 
flips must have had no idea about their initial state (although he could 
have known about a correlation between them). This suggests a connection 
between envariance and ignorance that will lead us to Born's rule.

In quantum theory envariance is possible for pure joint states. This is 
because of the nature of quantum correlations -- of
entanglement. 
%
Assume that the joint state $|\psi_{\cal SE}\rangle$ is pure, expressed in
the Schmidt form, Eq. (1) with complex coefficients $\alpha_k$ and
with $\{|\sigma_k\rangle \}$ and $\{ | \varepsilon_k \rangle \}$ orthonormal.
There exist transformations $u_{\cal S}$ acting on ${\cal S}$ alone that can be
cancelled by a acting on ${\cal E}$ alone for an arbitrary joint pure state of
two systems. Pairs of $u_{\cal S}$ and $u_{\cal E}$ that satisfy:
$$ u_{\cal S} \otimes u_{\cal E} |\psi_{\cal SE}\rangle \ = \
\sum_{k=1}^N \alpha_k (u_{\cal S}|\sigma_k \rangle )
( u_{\cal E} | \varepsilon_k \rangle  ) \ = \
\sum_{k=1}^N \alpha_k |\sigma_k \rangle | \varepsilon_k \rangle  $$
exist for an arbitrary set of coefficients $\alpha_k$: Such $u_{\cal S}$
are generated by the Hamiltonians of the system that have Schmidt eigenstates
$\{ |\sigma_k \rangle \}$. For, in this case, the only effect on the system
is the rotation of the phases of the coefficients in the Schmidt decomposition:
$$ u_{\cal S} |\sigma_k \rangle = e^{i \omega^{\cal S}_k t_{\cal S}}
|\sigma_k \rangle
= e^{i \varphi_k} |\sigma_k \rangle \ . \eqno(4)$$
Any such $u_{\cal S}$ can be countered by a $u_{\cal E}$:
$$ u_{\cal E} |\varepsilon_k \rangle = e^{- i \omega_k^{\cal E} t_{\cal E}}
|\varepsilon_k \rangle  = e^{ - i( \varphi_k + 2 \pi l_k)} |\varepsilon_k 
\rangle \ , \eqno(5)$$
where $l_k$ is an integer.
Phases of the coefficients in the Schmidt decomposition can be arbitrarily
changed by local interactions. Note that we are affecting phases of the
coefficients solely by acting on the states.
Note also that -- in this case -- eigenvalues of the system Hamiltonian
$\{ \omega_k^{\cal S} \}$ can be selected at random. It is the matching of
$e^{i \omega^{\cal S}_k t_{\cal S}}$ with $e^{-i\omega_k^{\cal E}t_{\cal E}} $
(allowing for the obvious freedom in choosing the eigenvalues of the 
Hamiltonian of the environment and of the associated times) that matters.

We conclude that an envariant description of the system alone must ignore
phases of the coefficients in Eq. (1). Such descrpition must be based on
a set of pairs $\{|\alpha_k|, |\sigma_k \rangle \}$. Hence, something with
the information content of the reduced density matrix (i.e., an object
dependent solely on $|\alpha_k|$ and on the associated states) provides
a complete description of ${\cal S}$ alone
given that the overall state of ${\cal SE}$ has a form we have assumed.
The same conclusion can be reached by appealing to causality: Phases of
Schmidt coefficients can be influenced by acting on ${\cal E}$ alone. If this
could be detected by measuring ${\cal S}$, faster than light communication
would become possible.

To justify Born's rule we still need
the relation between $|\alpha_k|$ and probabilities. On the other hand,
from the uniqueness of Schmidt decomposition we have already recovered 
the set of the preferred states $\{ |\sigma_k \rangle \}$: They are the
``fixed points'', eigenstates of the envariant transformations $u_{\cal S}$ 
(as well as of the local Hamiltonians that generate such 
$u_{\cal S}$). In a sense, this is yet another derivation of the 
pointer states$^{6}$. Schmidt states were known to enjoy an intimate
relationship with them, and have been even regarded as
``instantaneous pointer states''$^{14, 15}$.

The set $\{ |\sigma_k \rangle \} $ is not unique when the absolute values of
a subset of the coefficients $|\alpha_k|$ are equal and non-zero. I now turn
to investigate this case. This will lead us to Born's rule -- to the
relation between the coefficients $\alpha_k$ of the corresponding set
of the candidate pointer states $\{ |\sigma_k \rangle \} $ and their
probabilities. Entangled state vector:
$$ |\bar \psi_{\cal SE} \rangle \ = \ \sum_{k=1}^N |\alpha | e^{i \varphi_k}
|\sigma_k \rangle |\varepsilon_k \rangle \eqno(6)$$
with all the coefficients of equal magnitude has a much larger set of envariant
properties than Eq. (1): Now any orthonormal basis can be regarded as Schmidt. 
In particular, unitary transformation diagonal in a Hadamard transform of 
any pair of basis states of $\{ |\sigma_k \rangle \} $ generates a different 
looking transformation of ${\cal S}$ which consists of a sum of an identity 
and a ``swap'':
$$ u_{\cal S} (i \leftrightarrow j) = e^{ i \varphi_{i,j}} |\sigma_i \rangle
\langle \sigma_j | + h.c. \ . \eqno(7a)$$
Swap is envariant -- it can be undone by a ``counterswap'':
$$ u_{\cal E} (i \leftrightarrow j) = e^{ i (\varphi_{i,j} + \varphi_i
- \varphi_j + 2 \pi l_{ij})} |\varepsilon_i \rangle \langle \varepsilon_j | + h.c. \eqno(7b)$$
Above $l_{ij}$ is an integer. 

An $i \leftrightarrow j$ swap switches $|\sigma_i \rangle$ and
$|\sigma_j \rangle$. After the associated counterswap, 
also the states of ${\cal E}$ and their phases ``get swapped''.
Thus, iff $|\alpha_i| = |\alpha_j|$;
$$ (u_{\cal S}(i \leftrightarrow j) \otimes u_{\cal E} (i \leftrightarrow j))
|\bar \psi_{\cal SE} \rangle \ =  |\bar \psi_{\cal SE} \rangle \ , \eqno(8)$$
which proves envariance for $|\bar \psi_{\cal SE} \rangle $ of Eq. (6) under
swaps (and, more generally, under swaps of states that have same absolute 
values of Schmidt coefficients).

To connect envariance under swaps with probabilities we remark that all of
the states of the system described by Eq. (6) can be exchanged this
way leaving the overall state unchanged. This can make no observable 
difference to the state of the system ${\cal S}$ alone when it is 
perfectly entangled (Eq. (11)) with some "environment'': Joint state 
$|\bar \psi_{\cal SE}\rangle$ is envariant under swaps. 
When all of the coefficients of swapped states are equal, 
the observer with access to ${\cal S}$ alone cannot detect the effect 
of the swap. 

Let us now make a rather general (and a bit pedantic) 
assumption about the measuring process: When the states are swapped 
in the state vector $\psi_{\cal SE}$, corresponding probabilities get 
re-labeled ($i \leftrightarrow j$). This leads us to 
conclude immediately that the probabilities for any two swappable 
$|\sigma_k \rangle$ are equal. Moreover, when all of the orthonormal 
states with non-zero coefficients are swappable, and there are a total 
$N$ of them, probability $p_k = p (\sigma_k)$ of each must be:
$$ p_k \ = \ 1/N \ , \eqno(9a)$$
by normalisation (we assume that states that do not appear in Schmidt
decomposition have zero probability). Furthermore, probability of any subset 
with $n$ mutually exclusive (orthonormal) $\{| \sigma_k\rangle \}$ is:
$$ p_{k_1 \vee k_2 \vee \dots \vee k_n} = n / N \ . \eqno(9b)$$

This case with equal absolute values of the coefficients was straightforward. 
Consider now a general case with unequal coefficients.
To avoid cumbersome notation that may obscure key ideas we focus on
the case with only two non-zero coefficients:
$$ |\psi_{\cal SE} \rangle = \alpha |0\rangle |\varepsilon_0 \rangle +
\beta |1 \rangle |\varepsilon_1 \rangle \eqno(10a)$$
and assume that they can be written as:
$$ \alpha = e^{i \varphi_0} \sqrt {m/M}; \ \
\beta = e^{i \varphi_1} \sqrt {(M-m)/M}  \ . \eqno(10b)$$
When there are no $m$ and $M$ for which Eq. (10b) holds exactly, we can still
put upper and lower bounds on $|\alpha|$ and $|\beta|$ by taking a sequence
of increasing $M$ and $m_+=m_-+1$ such that $\sqrt{m_+/M} > |\alpha| 
> \sqrt {m_-/M}$ and, by continuity, recover our conclusions in the
$M \rightarrow \infty$ limit.

The strategy now is to convert the entangled state of Eq. (10) with unequal
coefficients into an entangled state with equal coefficients, and then to 
apply envariance-based reasoning that has led to Eq. (9). ``Fine-graining'' 
is a well - known trick, used on similar occassions in the classical 
probability textbooks, but applicable also in the quantum context$^{16,17}$. 
To implement it, we need {\it two} systems ${\cal C}$ and ${\cal E}$ 
correlated with the ``system of interest'' ${\cal S}$. One can think of 
${\cal C}$ (which is added as a ``counterweight'' to justify invariant 
swapping) and ${\cal E}$ (which plays the role of the ``real environment'', 
allowing us to disregard
phases for reasons discussed previously) as two parts of a single
``real environment''. There are a number of ways to motivate this split of
the original ${\cal E}$ into the counterweight ${\cal C}$ and the new
${\cal E}$. Amongst them we shall choose a shortcut and simply assert that
${\cal C}$ with the right attributes (to be listed below) interacts with
(pre-measures) ${\cal E}$ so that the joint state has a form:
$$ |\psi_{\cal SCE} \rangle \sim
   e^{i \varphi_0} \sqrt{m} |0\rangle |C_0\rangle | \varepsilon_0 \rangle \ +
\ e^{i\varphi_1} \sqrt {M-m} |1\rangle |C_1\rangle |\varepsilon_1\rangle \eqno(11)$$
where all the states are orthonormal. We assume that $|C_0\rangle$ and 
$|C_1\rangle$ can be expressed in a different orthonormal basis 
$\{ |c_j\rangle \}$:
$$ |C_0\rangle = \sum_{k=1}^m |c_k\rangle/\sqrt m ; \
|C_1\rangle = \sum_{k=m+1}^M |c_k\rangle/\sqrt {M-m} \eqno(12)$$
This requires that the relevant subspaces of ${\cal H}_{\cal C}$
correlated with ${\cal S}$ have sufficient dimensionality.

Envariance we now exploit is associated with the existence of counterswaps 
of ${\cal E}$ that undo swaps of the joint state of the composite
system ${\cal SC}$. To exhibit it, we let ${\cal C}$ interact with ${\cal E}$ 
again (e.g., by employing ${\cal C}$ as a control to carry out 
a {\tt c-shift}$^5$)
so that $|c_k\rangle |e_0 \rangle \rightarrow |c_k\rangle |e_k \rangle $,
where $|e_0\rangle$ is the initial state of ${\cal E}$
and $\langle e_k | e_l \rangle = \delta_{kl}$ in some suitable
orthonormal basis $\{|e_k\rangle \}$. Thus;
\begin{eqnarray*}
|\Psi_{\cal SCE}\rangle &\sim& e^{i \varphi_0} \sqrt{m} ~~ |0\rangle ~~ \sum_{k=1}^m  {{ |c_k\rangle |e_k\rangle } \over \sqrt m} \\  
&+& e^{i\varphi_1}\sqrt {M-m}~~|1\rangle \sum_{k=m+1}^M {{|c_k\rangle |e_k\rangle}
\over \sqrt{M-m}}\quad (\mathrm{13}a)
\end{eqnarray*}
obtains. This ${\cal CE}$ interaction can happen far from ${\cal S}$, so it
cannot influence probabilities observer will attribute to ${\cal S}$.
$|\Psi_{\cal SCE}\rangle $ is envariant under swaps of the states 
$|s, c_k\rangle$ of the composite ${\cal SC}$ system (where $s$ stands 
for 0 or 1, as needed) that are present (i.e., appear with a non-zero 
amplitude) in the $\Psi_{\cal SCE}$ above. This is made even more
apparent by carrying out the obvious cancellations;
$$ |\Psi_{\cal SCE}\rangle \sim e^{i \varphi_0} \ 
\sum_{k=1}^m |0, c_k\rangle |e_k\rangle +
e^{i\varphi_1} \sum_{k=m+1}^M |1, c_k\rangle |e_k\rangle . \eqno(13b)$$
We conclude (having repeated checks patterned on the obvious modifications
of Eqs. (7) and (8)) that $ p_{0,k} = p_{1,k} = 1/M $,
and, by virtue of Eq. (9b), probabilities of $|0\rangle $ and $|1\rangle$ are:
$$ p_0 = {m \over M} = |\alpha|^2; \ \ \
p_1 = {{M-m} \over M} = |\beta|^2 \eqno(14)$$
This is Born's rule. We have derived it from the most fundamental
properties of quantum physics, including in particular
these embodied in envariance. In contrast with the other
derivations, it relies on the most quantum of foundations -- the
incompatibility of the knowledge about the whole and about the parts, mandated 
by entanglement. It explains how Born's rule arises in a purely
quantum setting, i.e., without appeals to ``collapse'', ``measurement'', or
any other such {\it deus ex machina} imposition of the symptoms of classicality
that violate the unitary spirit of quantum theory. Generalisation to more than
two states of the system is straighforward.

It is also possible to establish the connection between the above approach
based on envariance and relative frequencies: Consider an ensemble of 
${\cal N}$ distinguishable ${\cal SCE}$ triplets all in the state
given by Eq. (11). The state of the ensemble is then;
$$|\Upsilon^{\cal N}_{\cal SCE}\rangle = \otimes_{\ell=1}^{\cal N}
|\psi_{\cal SCE}^{(\ell)}\rangle \eqno(15)$$  
We can now go through the steps, Eqs. (12)-(14) for each of the triplets,
and think of ${\cal C}$ as a counter, a detector in which
states $|c_1\rangle \dots |c_m\rangle$ of Eq. (13b) record ``0'' 
in ${\cal S}$, while $|c_{m+1}\rangle \dots |c_M\rangle$ record ``1''.
Carrying out the tensor product and counting terms with $n$ detections
of ``0'' yields the total:
$$ \nu_{\cal N}(n) = {{\cal N} \choose n} m^n (M-m)^{{\cal N} - n}
\eqno(16)$$
This leads immediately to the probability of $n$ 0's:
$$ p_{\cal N}(n) = {{\cal N} \choose n} |\alpha|^{2n} |\beta|^{2({\cal N}-n)}
\simeq { 
e^{- { 1 \over 2}\bigr({{n - |\alpha|{\cal N}} \over \sqrt {\cal N} |\alpha \beta|}\bigl)^2} 
\over {\sqrt{2 \pi {\cal N}} |\alpha \beta|}}
\eqno(17)$$
which in the limit of large ${\cal N}$ can be approximated by a Gaussian
with $\langle n \rangle = | \alpha|^2 {\cal N}$, establishing the desired 
link between relative frequency and Born's rule. Moreover, this strategy 
avoids circular use of scalar product (that invalidates$^{10,11}$ previous 
derivations based on Everett's ``Many Worlds'' framework$^{18-20}$).

Note that the steps involving the counterweight/counter ${\cal C}$ do not
need to be implemented -- our conclusions are based on the fact that they
{\it can} be implemented. As the implementation can happen in a region
causally disconnected from ${\cal S}$ anyway, the conclusions about
probabilities that emerge are based on the nature of quantum states of joint
systems (i.e., the fact that they inhabit tensor product of the respective
Hilbert spaces), and on the nature of the transformation that allow for
envariance. 

The setting (involving entanglement between ${\cal S}$ and ${\cal E}$) 
that has led to Born's rule is that of einselection and decoherence$^{3-7}$. 
Of course, as we have attempted to validate foundations
of decoherence, we have not relied on it. But the very fact that
Born's rule naturally obtains with the help of environment -- as does
decoherence -- seems to validate this view of the emergence of the classical.

This last remark requires some elaboration: One might have hoped to arrive
at the probability interpretation without appealing to the environment.
Indeed, there were many attempts in this vein$^{17-22}$ to both derive and 
interpert Born's rules. I do not see how attempts that do not obliterate 
phases in some manner (e.g., by invoking ``collapse'', as Refs. 2, 21 
and 22 do) can succeed in obtaining probabilities of $|\sigma_k\rangle$ in 
a pure state,
e.g.  $ |\chi_{\cal S} \rangle = \sum_{k=1}^N \alpha_k |\sigma_k \rangle $.
Equal probabilities for $|\sigma_k \rangle $ must imply that 
swapping of the alternatives in the state of the form:
$ |\bar \chi_{\cal S} \rangle =
\sum_{k=1}^N e^{i \varphi_k} |\alpha| ~ |\sigma_k \rangle $
should not be detectable if the obvious consequence (that ``all the potential 
outcomes are equivalent'') is to follow$^{17}$. But this is demonstrably 
{\it not} the case. For instance, states $|1\rangle + |2\rangle -|3\rangle$ 
and $|3\rangle + |2\rangle -|1\rangle$ are distinguishable through obvious 
interference measurements. Thus any swapping would be signalled by a change 
in relative phases, which in an isolated system are perfectly detectable. 

Envariance of entangled quantum states follows from the non-locality of 
states and from the locality  of systems, or, put a bit differently, from 
the coexistence of perfect knowledge of the whole and complete ignorance  
of the parts. This very quantum fact provides the basis of our derivation 
of Born's rule. We note that, while quantum theory and the presently 
available data (e.g., obtained in course of tests of Bell's inequalities) 
are consistent with envariance, its validity has not been deliberately 
verified. Such experiments would be fundamentaly important.

Stimulating discussions with Manny Knill are gratefully acknowledged.
This research was supported in part by National Security Agency.

%


\end{document}